# DERIVATION OF CONDITIONS FOR A BRANS-DICKE COUPLING CONSTANT OF ORDER UNITY BE CONSISTENT WITH SOLAR SYSTEM BOUNDS

Jean Paul Mbelek[1]

1. Sangha Center for Astronomy, Astrophysics and Cosmology, Sangha, Mali

Abstract : We provide proofs of some assumptions recently made by F. O. Minotti to conclude on the possibility that an additional scalar field minimally coupled to gravity may help to reconcile a Brans-Dicke coupling constant $\omega_0 \sim 1$ with solar system bounds.

Recently, F. O. Minotti [1] has used the KK-psi theory [2, 3, 4] to account for experimental results indicating the appearance of unusual forces on asymmetric electromagnetic resonant cavities. In the same study, the author found an interesting theoretical result for the KK-psi theory. Namely, the external scalar field, $\psi$, of the KK-psi theory could reconcile the solar system tests with a Brans-Dicke coupling constant $\omega_0 \sim 1$ [5]. To arrive at the latter conclusion, the author made two major assumptions about the order of magnitude of two partial derivatives $\partial\beta_{mat}/\partial\phi$ and $\partial\beta_{mat}/\partial\psi$ of the coupling function $\beta_{mat} = \beta_{mat}(\psi, \phi)$ of $\psi$ to matter, where $\phi$ denotes the genuine scalar field of the 5D Kaluza-Klein theory (KK). In the following, we show that the assumptions put forward by F. O. Minotti are actually direct consequences of the KK-psi theory itself. To start with, let us recall the equation of motion of a minimally coupled scalar field $\psi$ with self-interaction potential $U = U(\psi)$ and source term J, this reads

$$\nabla^\nu \nabla_\nu \psi = - \partial U/\partial \psi + J. \quad (1)$$

As one knows, the vacuum expectation value (vev), $\psi_0$, of the $\psi$-field can be derived from equation (1) above by setting the source term J to zero and requiring that $\psi_0$ still be a solution of equation (1). This procedure yields the appropriate equation to solve in order to estimate $\psi_0$, namely, $(\partial U/\partial \psi)(\psi_0) = 0$. In the same manner, we may apply the same procedure to the equations of motion of the scalar fields $\phi$ and $\psi$. The equations of the $\phi$ and $\psi$ fields read respectively,

$$(2\omega + 3)\nabla^\nu \nabla_\nu \phi = - (d\omega/d\phi) \nabla^\nu \phi \nabla_\nu \phi - 4\pi G_0 \varepsilon_0/c^2 \phi (d\lambda/d\phi) F_{\mu\nu} F^{\mu\nu} + 8\pi G_0/c^4 T^{mat}$$
$$+ \phi [½ \nabla^\nu \psi \nabla_\nu \psi - U(\psi) - J\psi] - (\partial J/\partial \phi)\psi\phi, \quad (2)$$

and

$$\nabla^\nu \nabla_\nu \psi + (\nabla^\nu \psi \nabla_\nu \phi)/\phi = - \partial U/\partial \psi - J - (\partial J/\partial \psi) \psi + (\beta/\phi) 8\pi G_0/c^4 T^{mat}, \quad (3)$$

where the source term of the $\psi$-field reads

$$J = \beta_{mat}(\psi, \phi) 8\pi G_0/c^4 T^{mat} + \beta_{EM}(\psi, \phi) 4\pi G_0 \varepsilon_0/c^2 F_{\mu\nu} F^{\mu\nu} + \beta_\phi(\psi, \phi) T^\phi. \quad (4)$$

Now, in the absence of the electromagnetic field, the equations (2) and (3) of the $\phi$ and $\psi$ fields should still be satisfied if these scalar fields are not excited. Thus, on account that $\beta_{mat}(\psi_0, \phi_0) = 0$, $U(\psi_0) = 0$ and $(\partial U/\partial \psi)(\psi_0) = 0$, replacing $\phi$ and $\psi$ by their respective vevs $\phi_0 = 1$ and $\psi_0$ in equations (2) and (3) yields,

$8\pi G_0/c^4 T^{mat} [1 - (\partial\beta_{mat}/\partial\phi)(\psi_0, \phi_0)\psi_0] = 0$, (5)

and

$8\pi G_0/c^4 T^{mat} [\beta - (\partial\beta_{mat}/\partial\psi)(\psi_0, \phi_0)\psi_0] = 0$. (6)

Since relations (5) and (6) should be fulfilled for any matter distribution, it follows that

$(\partial\beta_{mat}/\partial\phi)(\psi_0, \phi_0)\psi_0 = 1$, (7)

and

$(\partial\beta_{mat}/\partial\psi)(\psi_0, \phi_0)\psi_0 = \beta$. (8)

Relations (7) and (8) are conditions required for the Brans-Dicke coupling constant $\omega_0$ be close to unity while satisfying the solar system constraints at the same time.

References

1. F. O. Minotti, (2013), arXiv:1302.5690v2 [gr-qc]

2. J. P. Mbelek and M. Lachièze-Rey, (2002), Grav. & Cosmol. 8, 331.

3. J. P. Mbelek and M. Lachièze-Rey, (2003), Astron. Astrophys. 397, 803.

4. J. P. Mbelek, (2004), Astron. Astrophys. 424, 761.

5. C. H. Brans and R. H. Dicke, (1961), Phys. Rev. 124, 925.